\documentclass[aps,prb,twocolumn,floatfix,bibliography,superscriptaddress]{revtex4}
\usepackage{graphics,bm,amsmath,amssymb,natbib,url,epsfig,psfrag,xcolor}
\usepackage{graphicx}
\usepackage{subfigure}
\usepackage[normalem]{ulem}

\usepackage[mono]{inconsolata}
\usepackage{xspace}

\usepackage[unicode=true,pdfusetitle,
bookmarks=true,bookmarksnumbered=false,bookmarksopen=false,
breaklinks=false,pdfborder={0 0 0},pdfborderstyle={},backref=false,colorlinks=true]
{hyperref}

\hypersetup{colorlinks,allcolors=blue}

\begin{document}
	\newcommand{\Tr}{\text{Tr}}
	\newcommand{\beq}{\begin{equation}}
		\newcommand{\eeq}{\end{equation}}
	\newcommand{\eran}[1]{{\color{blue}#1}}
	
	\newcommand{\be}{\begin{equation}}
		\newcommand{\ee}{\end{equation}}
	\newcommand{\bea}{\begin{eqnarray}}
		\newcommand{\eea}{\end{eqnarray}}
	\newcommand{\ES}[1]{{\color{violet}#1}}
	\newcommand{\sarath}[1]{{\color{red}#1}}
	\newcommand{\YM}[1]{{\color{blue}#1}}
	\newcommand{\JF}[1]{{\color{orange}#1}}
	\def\bs#1\es{\begin{split}#1\end{split}}	\def\bal#1\eal{\begin{align}#1\end{align}}
	\newcommand{\nn}{\nonumber}
	\newcommand{\sgn}{\text{sgn}}
	
	\title{Direct signatures of Anderson orthogonality catastrophe in nonequilibrium quantum dots}

	\author{Sarath Sankar}
	\affiliation{School of Physics and Astronomy, Tel Aviv University, Tel Aviv 6997801, Israel}
	\affiliation{Department of Physics, Ben-Gurion University of the Negev, Beer-Sheva 84105, Israel}
	\author{Joshua Folk}
	\affiliation{Quantum Matter Institute, University of British Columbia, Vancouver, British Columbia, Canada}
	\affiliation{Department of Physics and Astronomy, University of British Columbia, Vancouver, British
		Columbia, Canada}
	\author{Yigal Meir}
	\affiliation{Department of Physics, Ben-Gurion University of the Negev, Beer-Sheva 84105, Israel}
	\author{Eran Sela}
	\affiliation{School of Physics and Astronomy, Tel Aviv University, Tel Aviv 6997801, Israel}

	\date{\today}
	\begin{abstract}
		We propose schemes for  unambiguous direct observation of Anderson orthogonality catastrophe (AOC) effects in a quantum dot coupled to a charge detector, enabling experimental estimation of the AOC exponent $\alpha$. We show that certain easy-to-measure observables have a robust dependence on $\alpha$ in the non-equilibrium regimes of source-drain voltage bias or thermal imbalance. Our results are obtained using a rate equation formalism in which the AOC effects on tunnel rates are incorporated in an exact manner, and directly support recent experimental results.
	\end{abstract}
	\maketitle
	\section{Introduction}
	The Anderson orthogonality catastrophe (AOC) occurs due to the response of a Fermi sea to a sudden change in  local potential~\cite{anderson1967infrared}. It is characterized by an exponent $\alpha$ describing the power-law decay of the overlap of the perturbed and unperturbed many-body ground state in the thermodynamic limit. First proposed over 50 years ago in order to explain features of X-ray photoemission in metals~\cite{nozieres1969singularities}, the AOC has since been shown to be relevant in systems as diverse as semiconductor quantum wells~\cite{skolnick1987observation}, ultra-cold atomic gases\cite{PhysRevX.2.0410201}, and self-assembled quantum dots\cite{benedict1998fermi,hapke2000magnetic}.
	
	The essential ingredients of the AOC can be conveniently isolated and controlled by building them up piece by piece in one of the simplest possible mesoscopic circuits: a quantum dot (QD) capacitively coupled to a charge detector. The detector Fermi sea is affected by the charge state of the QD and AOC arises when  tunnel events occur in the QD.   In this arrangement, the AOC in the detector manifests as measurement  backaction~\cite{aleiner1997dephasing,sankar2024detector,sankar2025back} on the physical observables of the QD, such as the current. When strong enough, AOC can drive phase transitions~\cite{goldstein2010population,ma2023identifying,ma2025engineering}. 
	The AOC exponent in general also receives additional contributions from  interactions of the QD with its lead~\cite{borda2007theory,krahenmann2017fermi} as in the interacting resonant level model~\cite{Ma_2025}, with other localized levels~\cite{kashcheyevs2009quantum}, or with other nearby gates.
	
	In fact, the connection between measurement backaction and AOC was identified nearly thirty years ago, in an analysis of how charge measurement can affect a which-path interferometry experiment.\cite{aleiner1997dephasing}  Although the primary backaction effect of charge measurement is often the dephasing induced by shot noise in the strongly-biased detector \cite{buks1998dephasing,avinun2004controlled,kung2009noise,bischoff2015measurement,ferguson2023measurement}, the authors of Ref.~\onlinecite{aleiner1997dephasing} pointed out that even with zero bias applied to the detector, AOC backaction remains as a fundamental limit on coherent dynamics in a measured quantum system.  However, an unambiguous experimental observation of AOC effects, including the determination of $\alpha$, in QDs coupled to detectors remains elusive, in part because the experimental observables that would clearly distinguish the effect were not well understood. 
	
\begin{figure}
		\includegraphics[width=\columnwidth]{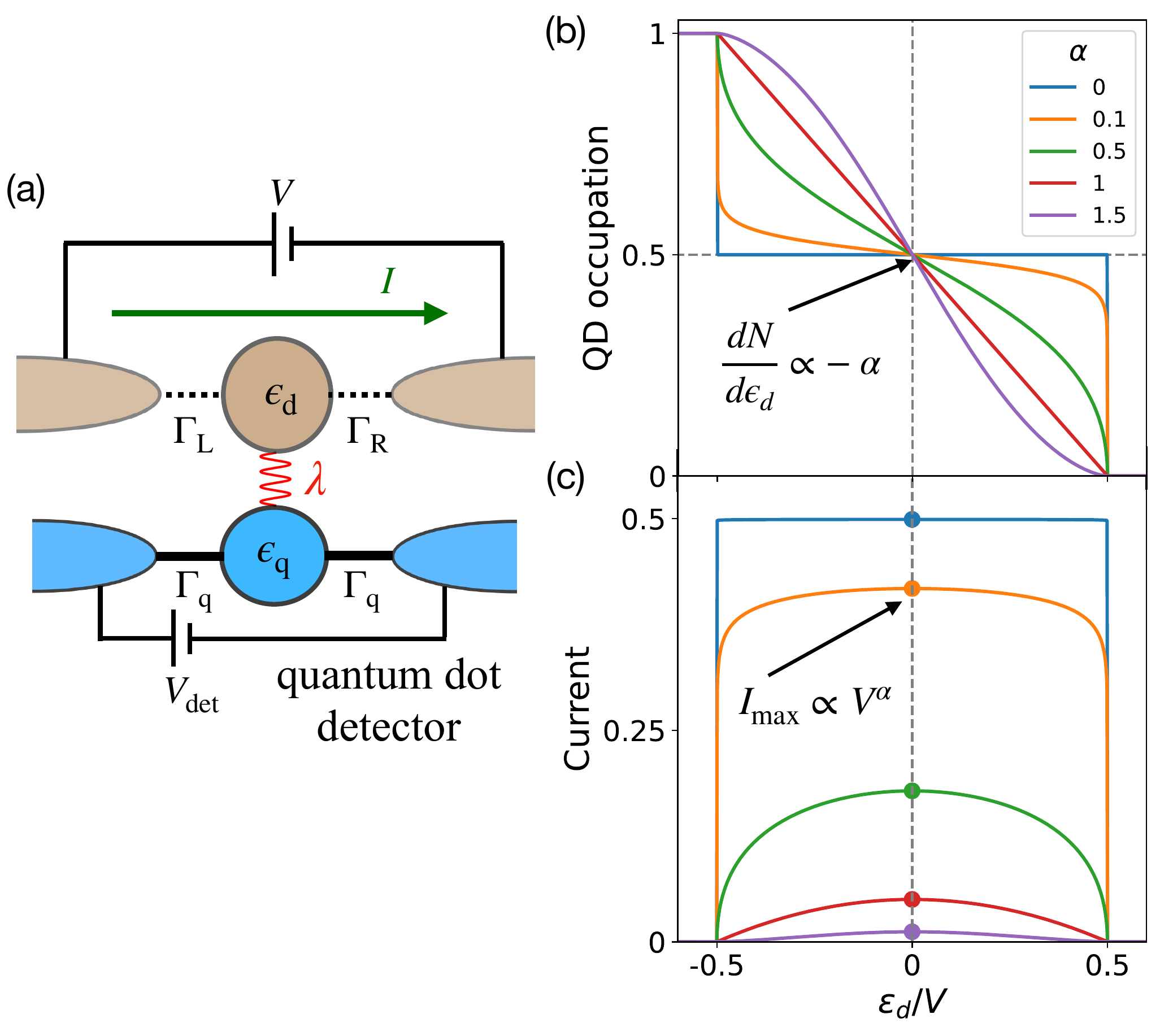}
		\caption{(a) Schematic of a QD  weakly tunnel coupled $(\Gamma_{{\rm{L,R}}})$ to voltage biased leads, and capacitively coupled $(\lambda)$ to a quantum dot detector. $\alpha$ is a dimensionless exponent associated with the Anderson orthogonality catastrophe in the detector due to tunnel events in the QD.  Panels (b), (c) discuss the spinless case of the QD with $\Gamma_{\rm{L}}=\Gamma_{\rm{R}}=\Gamma$ and  the limit $T,\,V_{\rm{det}}\to 0$. 
			(b) Average occupation of the QD, $N(\epsilon_d)$, and the average steady state current, $I(\epsilon_d)$, depends strongly  on $\alpha$ in the energy window $\epsilon_d \in [-V/2,V/2]$ as shown respectively in (b) and (c).
			The charge susceptibility in (b)  obeys $\partial N/\partial \epsilon_d |_{\epsilon_d=0} \propto -\alpha$, and provides a direct measure of $\alpha$, see Eq.~\eqref{eq:alpha_suscep_spinless}.
			The current peak (dots) in (c) that occurs at $\epsilon_d=0$ is proportional to $V^\alpha$ and provides another direct measure of $\alpha$, see Eq.~\eqref{eq:alpha_cur_spinless}.
		}
		\label{fig:1}
	\end{figure}

	Here, within the regime of a weakly tunnel coupled QD that is driven strongly out of equilibrium, we describe two routes to isolate AOC backaction signatures, controlled by $\alpha$,  from other back action effects. We consider two sources of non-equilibrium: (i) source-drain voltage $V$ across the QD leads, see Fig.~\ref{fig:1}(a), and (ii) different temperatures in the leads of the QD and of the detector, see Fig.~\ref{fig:4}(a). In both  cases we identify direct approaches to experimentally measure $\alpha$ from the back action signatures on the QD observables such as its charge curve  or steady state current. The theoretical predictions presented in this paper directly support recent experimental results~\cite{grant2026aoc}.
	
	The rest of the paper is organized as follows. In Sec.~\ref{se:model} we present the model and the  correlation functions $A^\pm(E)$ encapsulating  AOC physics. In Sec.~\ref{sec:biased_QD} we describe AOC signatures on the current and occupation of the QD due to a nonequilibrium effect arising from a finite voltage bias. In Sec.~\ref{se:FES} we describe the additional effect of Fermi edge singularity in the system's leads. In Sec.~\ref{sec:dif_temp} we study AOC effects due to  thermal imbalance between the system and the detector.   We conclude in Sec.~\ref{se:con}.

	\section{Model}
	\label{se:model}
	We consider the model introduced in Ref.~\onlinecite{sankar2024detector}, with $H=H_{\rm{{sys}}}+H_{{\rm{{det}}},n}$, describing a system coupled to a detector. The system consists of a spinless QD (for a discussion of the spinful case see appendix \ref{sec:spin}) with energy level $\epsilon_d$ tunnel coupled to L and R leads  with tunnel coupling $\gamma_{{\rm{L/R}}}$,
	\begin{equation}
		\label{eq:H_sys}
		H_{\rm{{sys}}}=\epsilon_d d^\dagger d + \sum_{\substack{i={\rm{L,R}} \\ k}  }\epsilon_{k} c_{ik}^\dagger c_{ik} + \sum_{i={\rm{L,R}}}(\gamma_i c_{i1}^\dagger d+ {\rm{H. c.}}),
	\end{equation}
	where $d^\dagger$ creates an electron in the QD,  $c_{ik}^\dagger$ creates an electron with momentum $k$ and energy $\epsilon_k$ in  lead $i$, and $c_{i1}=\sum_k c_{ik}$. The tunneling linewidth is $\Gamma_{i}=2\pi \nu_0 \gamma_i^2$, where $\nu_0$ is the density of states in the leads. We consider weak tunneling in the QD, such that the temperature $T$ exceeds the tunneling linewidth  $T \gg \Gamma_{\rm{L},\rm{R}} $; in this regime quantum hybridization effects can be neglected.   
	In the spinless case there are only two charge states of the QD denoted $n=0,1$ corresponding to empty and occupied $d-$level. While $n$ denotes the eigenvalues of $d^\dagger d$, its average is denoted by $N=\langle d^\dagger d \rangle$. The absence of quantum hybridization effects in the considered high temperature regime implies only a probabilistic thermal superposition of these two charge states as described by a rate equation, see Eq.~(\ref{eq:rate_equation}) below.
	
	The second term in the Hamiltonian describes a quantum dot detector (QDD). Its Hamiltonian includes its coupling with the system's QD thorugh its dependence on $n$, 
	\begin{align}
		\label{eq:model}
		H_{{\rm{{det}}},n}&= \epsilon_{{\rm{q}}} f^\dagger f +\sum_{\substack{j={\rm{L,R}}\\ k}} (\epsilon_k \varphi_{j k}^\dagger \varphi_{j k} + v_j \varphi_{j k}^\dagger f +h.c. )  \nonumber\\&+\lambda \left(f^\dagger f-\frac{1}{2}\right)\left(n-\frac{1}{2}\right),
	\end{align}
	where $f^\dagger$ creates an electron in the QDD with energy level $\epsilon_q$, $\varphi_{j k}^\dagger$ creates an electron with momentum $k$ in the detector lead $j$ having density of states $\nu_d$, and $\lambda$ denotes the coupling strength between the system and the detector.
	The total tunneling width of the QDD is $\Gamma_{\rm{q}}=2\pi \nu_d (v_{\rm{L}}^2+v_{\rm{R}}^2)$. In contrast to the system QD, the  QDD is assumed to be strongly tunnel coupled to its leads, $\Gamma_{\rm{q}}\gg T$.
	
	In this model we control the AOC exponent $\alpha$ via  $\lambda$ and $\epsilon_{\rm{{q}}}$, as 
	$\alpha=(\delta/\pi)^2$ where,
	\begin{equation}
		\label{eq:phase_shift}
		\delta = \arctan\left(\frac{\epsilon_{{\rm{q}}}+\lambda/2 }{\Gamma_{{\rm{q}}}}\right)-\arctan\left(\frac{\epsilon_{{\rm{q}}}-\lambda/2 }{\Gamma_{ {\rm{q}} }}\right).
	\end{equation}
	\subsection{Tunneling rates}
	As discussed in Ref.~\onlinecite{sankar2024detector}, in the presence of the capacitive coupling to the detector, the tunnel rates of the system QD from/to the lead $i={\rm{L/R}}$ are given by
	\begin{eqnarray}
		\label{eq:tunin}
		\Gamma_{i}^{{\rm{in}}}&=&\Gamma_i \int dE\,f(E-\mu_i) A^+(E-\epsilon_d),\\
		\label{eq:tunout}
		\Gamma_{i}^{{\rm{out}}}&=&\Gamma_i \int dE\,(1-f(E-\mu_i)) A^-(E-\epsilon_d).
	\end{eqnarray}
	Here $f(E)=\frac{1}{e^{E/T}+1}$ is the Fermi function  and $A^{\pm}(E) = \int dt\,  e^{iEt} \tilde{A}^{\pm}(t)$
	where 
	\begin{equation}
		\label{eq:a_corr_time}
		\tilde{A}^{\text{\tiny +/}\text{\small -}}(t)={\rm{Tr}}\left[\rho_{0/1}e^{itH_{\rm{{det}},0}}e^{-itH_{\rm{{det}},1}}\right],
	\end{equation}
	with $\rho_n \propto e^{- H_{{{\rm{{det}}}},n}/T} $ denoting the detector density matrix for QD charge $n$.
	The functions $A^{\pm}(E)$ are the key objects that incorporate the AOC. They respectively denote the probability  that in the system QD tunnel-in/out process, the detector will absorb/emit an energy $E$. 
	
	We evaluate these functions numerically using the fact that for a fixed  $n$ the detector Hamiltonians $H_{\rm{{det}},n}$ are quadratic. For further details see Ref.~\onlinecite{sankar2024detector}. When the density of states of the detector is constant, methods such as bosonization can be applied to provide analytic expressions for these functions. 
	Particularly when the detector is at equilibrium, namely for $V_{{\rm{det}}}=0$, we have~\cite{aleiner1997dephasing}
	\begin{equation}
		\label{eq:a_pm_long_time}
		\tilde{A}^{\pm}(t)=\left(\frac{\pi T}{{\pm} i D \sinh(\pi T t) }\right)^{\alpha},
	\end{equation}
	where $D$ denotes a high energy cutoff below which the assumption of a constant density of state is valid. This formula applies at long times exceeding $1/D$.  For our QDD, the high energy cutoff  is associated with the detector bandwidth, $D\sim \Gamma_{\rm{q}}$. 
	The corresponding Fourier transforms are
	\begin{equation}
		\label{eq:apm_analytical}
		A^{\pm}(E)= \mathcal{A} \Gamma(1-\alpha) T^{\alpha-1}{\rm{Re}}\left[\frac{e^{{\pm} i\pi\alpha/2}\Gamma\left(\frac{iE}{2\pi T}+\frac{\alpha}{2}\right)}{\Gamma\left(1+\frac{iE}{2\pi T}-\frac{\alpha}{2}\right)}\right],
	\end{equation}
	where $\Gamma( \cdot )$ denotes the Gamma function and $\mathcal{A}$ is a nonuniversal prefactor depending on the high energy cutoff~\cite{sankar2024detector}. At zero temperature,
	\begin{equation}
		\label{eq:AT0}
		A^{\pm}(E)=\mathcal{A} \theta(\pm E) |E|^{\alpha-1}.
	\end{equation}
	Both Eqs.~(\ref{eq:apm_analytical}) and (\ref{eq:AT0}) are restricted to low energies, $E,T \ll D$.

	\section{AOC signatures in a QD with  voltage biased leads \label{sec:biased_QD}}
	Here, we describe AOC signatures that can be isolated when a non-equilibrium condition is generated by a large source-drain voltage bias applied across the QD, with the leads held at chemical potentials $\mu_{L/R}=\pm eV/2$, see Fig.~\ref{fig:1}(a). Although the system is driven out of equilibrium, the detector is assumed to be  at equilibrium, i.e. $V_{\rm{det}}\to 0$.  We now identify direct manifestations of $\alpha$ in the non-equilibrium behavior of the QD when $V \gg T$. 
	
	\subsection{Average occupation}
	The average QD occupation $N$ is governed  by the  rate equation
	\begin{equation}
		\label{eq:rate_equation}   \frac{dN}{dt}=(1-N) [\Gamma_L^{{\rm{in}}}+\Gamma_R^{{\rm{in}}}]-N [\Gamma_L^{{\rm{out}}}+\Gamma_R^{{\rm{out}}}].
	\end{equation}
	At steady state we have
	\begin{equation}
		\label{eq:occ_spinless_expr}
		N=\frac{\Gamma_L^{{\rm{in}}}+\Gamma_R^{{\rm{in}}}}{\Gamma_L^{{\rm{in}}}+\Gamma_R^{{\rm{in}}}+\Gamma_L^{{\rm{out}}}+\Gamma_R^{{\rm{out}}}}.
	\end{equation}
	At equilibrium, $V=V_{\rm{det}}=0$, the average occupation becomes a Fermi function, $N=f(\epsilon_d-\Delta)$, with $\Delta$ being a constant energy shift independent of $\epsilon_d$. This follows from the relation
	\begin{equation}
		\label{eq:apm_fdr}
		A^-(E)=e^{- (E +\Delta)/T}A^+(E),
	\end{equation}
	where $\Delta=F_{{\rm{det}}}(1)-F_{{\rm{det}}}(0)$, with $F_{{\rm{det}}}(n)$ denoting the free energy of the detector for the system QD charge state $n$. 
	Thus the charge curve, $N$ versus $\epsilon_d$, is not affected by $\alpha$ when $V=0$ except for the effective shift of $\epsilon_d$ which we ignore from here on.
	
	Now let us discuss the case $V\ne 0$ but consider the limit $T,\,V_{\rm{det}}\to  0$. In this limit from Eqs.~(\ref{eq:AT0}) we have  
	\begin{eqnarray}
		\label{eq:tunin_t0}
		\Gamma_{i}^{{\rm{in}}}(\epsilon_d)&\overset{T\to 0}{=}&\theta(\mu_i-\epsilon_d)\frac{\mathcal{A}\Gamma_i}{\alpha} (\mu_i-\epsilon_d)^\alpha,\\
		\label{eq:tunout_t0}
		\Gamma_{i}^{{\rm{out}}}(\epsilon_d)&\overset{T\to 0}{=}&\theta(\epsilon_d-\mu_i)\frac{\mathcal{A} \Gamma_i}{\alpha} (\epsilon_d-\mu_i)^\alpha.
	\end{eqnarray}
	Thus the tunnel rates show a power-law dependence on the difference of the non-equilibrium chemical potentials $\mu_{L/R}$ from the level $\epsilon_d$.
	Plugging these tunnel rates in Eq.~\eqref{eq:occ_spinless_expr} with $\mu_{L/R}=\pm V/2$ (setting $e=1$ here and in the following)) yields, $N(\epsilon_d<-V/2)=1$, $N(\epsilon_d>V/2)=0$, and for $V/2<\epsilon_d<V/2$ we have $N=\frac{\Gamma_L^{\rm{in}}}{\Gamma_L^{\rm{in}}+\Gamma_R^{\rm{out}}}$, giving
	\begin{equation}
		\label{eq:occ_t0}
		N \overset{T\to 0}{=} \frac{\Gamma_L (\frac{V}{2}-\epsilon_d)^\alpha}{\Gamma_L (\frac{V}{2}-\epsilon_d)^\alpha+ \Gamma_R (\frac{V}{2}+\epsilon_d)^\alpha}~~~\left(|\epsilon_d|<\frac{V}{2} \right).
	\end{equation}
	This is plotted in Fig.~\ref{fig:1}(b) for different values of $\alpha$. The charge plateau for $\alpha=0$ that occurs for $-V/2 <\epsilon_d < V/2$ is clearly  modified by a finite $\alpha$. Interestingly, the average occupation at the center of the step $\epsilon_d=\frac{\mu_L+\mu_R}{2}=0$, for any asymmetry $\Gamma_L/\Gamma_R$, is not affected by $\alpha$,
	\begin{equation}
		\label{eq:nc}
		N_c \equiv N(\epsilon_d=0)=\frac{\Gamma_L}{\Gamma_L+\Gamma_R},
	\end{equation}
	while the slope, i.e. charge susceptibility at that point, is given by
	\begin{equation}
		\left(\frac{\partial N}{\partial \epsilon_d}\right)_{\epsilon_d=0}=\frac{-4\alpha \Gamma_L \Gamma_R}{V(\Gamma_L+\Gamma_R)^2}.
	\end{equation}
	Using Eq.~\eqref{eq:nc} we then obtain a simple relation connecting the charge susceptibility to $\alpha$,
	\begin{equation}
		\label{eq:alpha_suscep_spinless}
		\alpha =\frac{-V}{4N_c(1-N_c)}\left(\frac{\partial N}{\partial \epsilon_d}\right)_{\epsilon_d=0}.
	\end{equation}
	This relation is convenient for experimentally inferring $\alpha$ since it is expressed in terms of quantities that are easy to measure and does not require the knowledge of parameters like $\Gamma_{L,R}$ or the lever arm.

	\begin{figure}
		\includegraphics[width=\columnwidth]{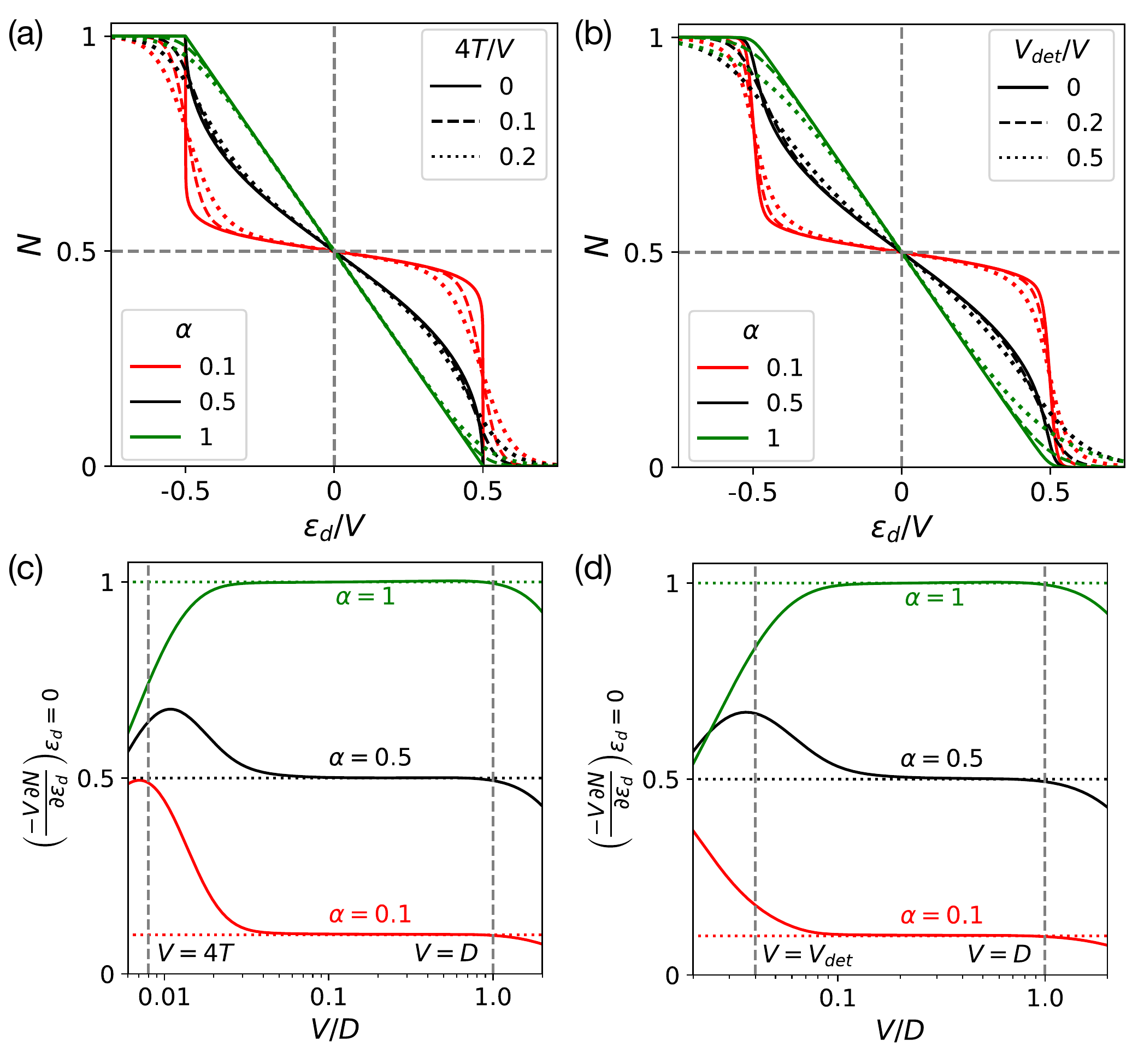}
		\caption{\label{fig:rob_occ_slope}  Effect of (a)  finite temperature and (b) detector bias  on the average occupation of the QD in the presence of $\alpha$. Here  $\Gamma_{\rm{L}}=\Gamma_{\rm{R}}$. For $V\gg T, V_{{\rm{det}}}$ the two charge steps get modified, however the middle region ($\epsilon_d \approx 0$) is nearly unaffected. (c,d): The charge susceptibility at $\epsilon_d=0$  remains a robust measure of $\alpha$ even in the presence of temperature and detector bias in the regime $T,\, V_{{\rm{det}}} \ll V < D$.
		} 
	\end{figure}
	
	\begin{figure}
		\includegraphics[width=\columnwidth]{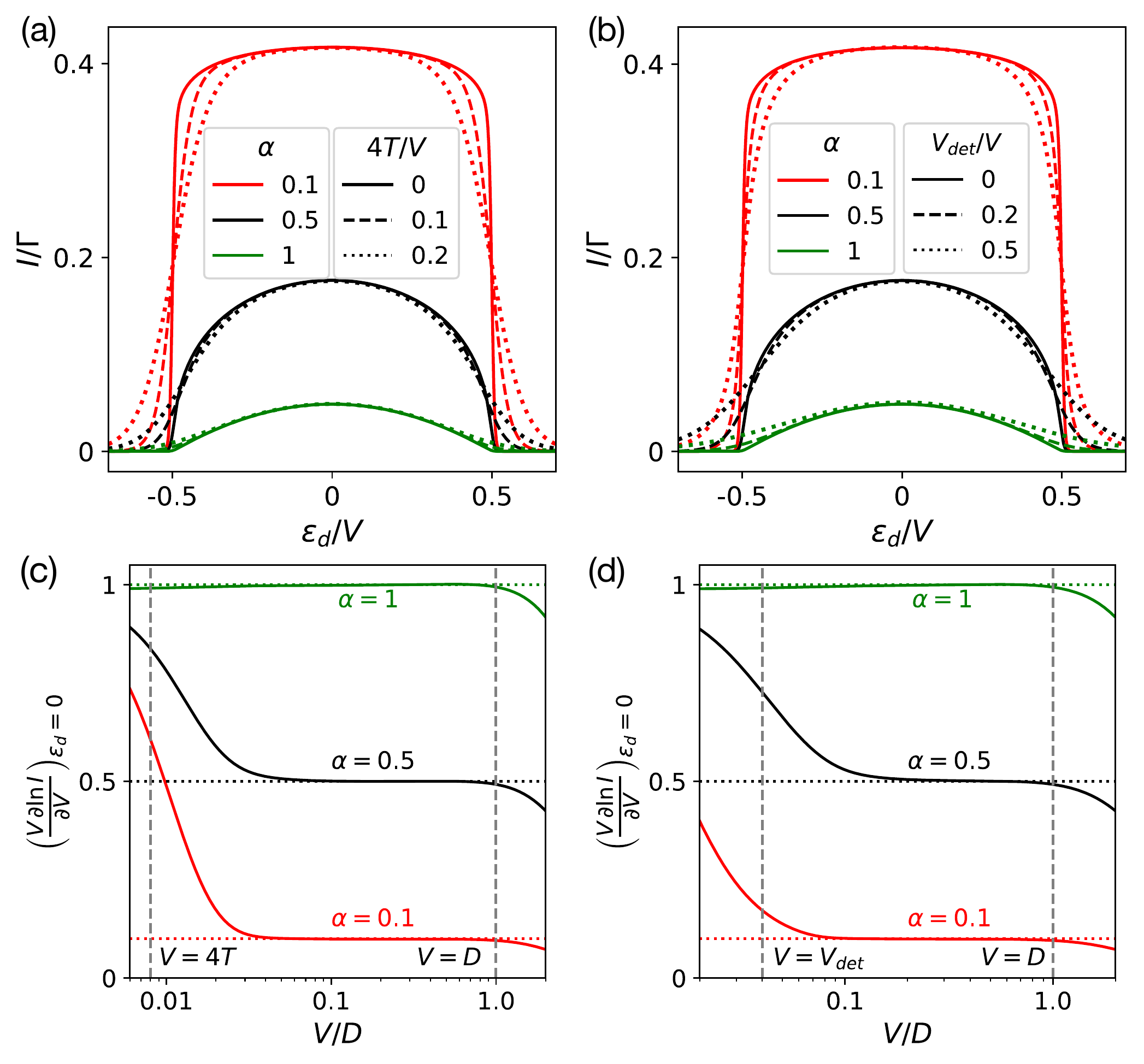}
		\caption{\label{fig:rob_cur_bias} Same as Fig.~\ref{fig:rob_occ_slope} but for the steady state current $I$. The voltage dependence of the  current peak at $\epsilon_d=0$ remains a robust measure of $\alpha$ in the regime $T,\, V_{{\rm{det}}} \ll V < D$.
		} 
	\end{figure}
	
	\subsection{Steady state current}
	The steady state current from $L$ to $R$ is given (as either evaluated in the left or right tunnel junction),
	\begin{equation}
		\label{eq:j_spinless_expr}
		I=(1-N)\Gamma_L^{{\rm{in}}}-N \Gamma_L^{{\rm{out}}}=N\Gamma_R^{{\rm{out}}}-(1-N) \Gamma_R^{{\rm{in}}}.
	\end{equation}
	Plugging the expression for tunnel rates in Eqs.~\eqref{eq:tunin_t0}, \eqref{eq:tunout_t0} and for $N$ in Eq.~\eqref{eq:occ_t0}, we first see that $I(\epsilon_d<\mu_R)=I(\epsilon_d>\mu_L)=0$. (Note that we focus only on the sequential tunneling current since we are actually interested in the regime $\epsilon_d \sim \frac{\mu_L+\mu_R}{2}$ and ignore the cotunneling current).  Next, for the QD level in-between the two chemical potentials, we get,
	\begin{equation}
		\label{eq:cur_t0}
		I(\epsilon_d)\overset{T\to 0}{=}\frac{\frac{\mathcal{A}}{\alpha}\Gamma_L\Gamma_R \left( \left(\frac{V}{2}\right)^2-\epsilon_d^2 \right)^\alpha}{\Gamma_L(\frac{V}{2}-\epsilon_d)^\alpha+\Gamma_R(\frac{V}{2}+\epsilon_d)^\alpha},~\left(|\epsilon_d|<\frac{V}{2}\right).
	\end{equation}
	This is plotted in Fig.~\ref{fig:1}(c) for different values of $\alpha$. The current peaks at the mid-point between the two chemical potentials i.e. at $\epsilon_d=0$ (independent of the asymmetry $\Gamma_L/\Gamma_R$), where it is given by
	\begin{equation}
		I(\epsilon_d=0)\overset{T\to 0}{=} \frac{\mathcal{A}}{\alpha \, 2^{\alpha}} \frac{\Gamma_L\Gamma_R}{\Gamma_L+\Gamma_R} V^\alpha.
	\end{equation}
	We clarify that this is derived in the sequential tunneling limit of small $\Gamma_{L,R}$, and indeed  in this limit when $\alpha=0$ the current for $\frac{V}{2}>|\epsilon_d|$  becomes independent of $V$ as $T\to 0$. When $\alpha$ is finite, the current shows a power law dependence on $V$:  $I\propto V^\alpha$.
	Such power laws in the I-V characteristics appeared for example  in transport through quantum point contacts in the fractional quantum Hall regime or in Luttinger liquids~\cite{wen2003quantum}; here the exponent originates from the AOC. Thus $\alpha$ can be inferred from the $V$ dependence of the peak,
	\begin{equation}
		\label{eq:alpha_cur_spinless}
		\alpha =\frac{\partial \ln[I(\epsilon_d=0)]}{\partial \ln V}.
	\end{equation}
	This relation is convenient for experimentally inferring $\alpha$ as it only requires measuring the $V$- dependence of the peak current. 
	
	Equations ~\eqref{eq:alpha_suscep_spinless} and ~\eqref{eq:alpha_cur_spinless} that relate  $\alpha$ to two different observables of the system QD are the main results of this paper. We recall  that these equations were derived under the assumptions that (i) the detector is at thermal equilibrium $(V_{\rm{det}}=0)$ and (ii) $T \to 0$.

	\subsection{Robustness at finite $T$ and $V_{\rm{det}}$}
	In this subsection we show that Eq.~\eqref{eq:alpha_suscep_spinless} and Eq.~\eqref{eq:alpha_cur_spinless} remain valid in the presence of a finite detector bias, $V_{\rm{det}}$ and finite $T$, provided that
	\begin{equation}
		\label{eq:valid_regime}
		\{T,V_{{\rm{det}}}\}\ll V < D.
	\end{equation}
	Note that a finite detector bias is required for experimentally measuring the QD occupation. 
	
	We follow Ref.~\onlinecite{sankar2024detector} and compute the functions $A^{\pm}(E)$  at finite $T$ and $V_{\rm{det}}$ using an exact numerical method generalizing the Nozier\'es-De Dominicis's solution~\cite{nozieres1969singularities}.
	Fig.~\ref{fig:rob_occ_slope} (a), (b) respectively shows the effect of finite $T$ and $V_{\rm{det}}$ on $N(\epsilon_d)$. 
	Their counterparts for $I(\epsilon_d)$ are shown respectively in Fig.~\ref{fig:rob_cur_bias}(a), (b).  While we display in these figures only the value of $\alpha$, in our calculation, the detector parameters are  set at  $\epsilon_q=0$, $\Gamma_q=1$ and $\lambda$ is varied to obtain different $\alpha$. We remark that in the regime given by Eq.~\eqref{eq:valid_regime}, the results in Figures ~\ref{fig:rob_occ_slope} and ~\ref{fig:rob_cur_bias} depend only on $\alpha$ and not on the specific detector parameters.

	Let us focus on Fig.~\ref{fig:rob_occ_slope}(a).
	Consider first the two steps of the charge curve at $\epsilon_d = \mu_{L/R}$. These steps display power-law behavior, for example near $\epsilon_d \to V/2$ we have at $T=V_{\rm{det}}=0$ that $N \sim (V/2-\epsilon_d)^\alpha$. We can see that these power law steps are thermally smeared by finite $T$. On the other hand, as long as $T \ll V$, the behavior near the middle of the plateau $\epsilon_d=0$ remains nearly unaffected by temperature and retains its power-law behavior $dN/dT \propto - \alpha$. As shown in Fig.~\ref{fig:rob_occ_slope}(c), this allows to extract $\alpha$ using Eq.~\eqref{eq:alpha_suscep_spinless} in the regime given by Eq.~\eqref{eq:valid_regime}. The high $V$ limitation in Eq.~\eqref{eq:valid_regime} stems from the fact that the power-law regime of the detector functions $A^{\pm}(E)$ is limited to a finite energy range dictated by the detector band-width $D=\Gamma_q$.
	
	The dependence on the detector bias is more involved, since in addition to supplying energy to the system, the AOC itself is modified in the non-equilibrium detector when there are multiple coupled Fermi edges~\cite{ng1996fermi,muzykantskii2003fermi,abanin2005fermi}. Nevertheless, a somewhat similar behavior occurs at finite $V_{\rm{det}}$ as compared to finite $T$, see Fig.~\ref{fig:rob_occ_slope}(b). Like in the finite $T$ case, we can apply Eq.~\eqref{eq:alpha_suscep_spinless} in the regime Eq.~\eqref{eq:valid_regime} to extract $\alpha$ as shown in Fig.~\ref{fig:rob_occ_slope}(d). 
	
	The above considerations also apply for the steady state current as shown in Fig.~\ref{fig:rob_cur_bias}. Specifically, Fig.~\ref{fig:rob_cur_bias}(a), (b)  show that in the regime Eq.~\eqref{eq:valid_regime}, the current peak is nearly unaffected due to a finite $T$ or $V_{\rm{det}}$; $\alpha$ can then be reliably extracted using  Eq.~\eqref{eq:alpha_cur_spinless} as shown in Fig.~\ref{fig:rob_cur_bias}(c), (d). 
	
	\subsection{Effects of tuning $\alpha$ by  $\epsilon_q$}
	
	\begin{figure}
		\includegraphics[width=\columnwidth]{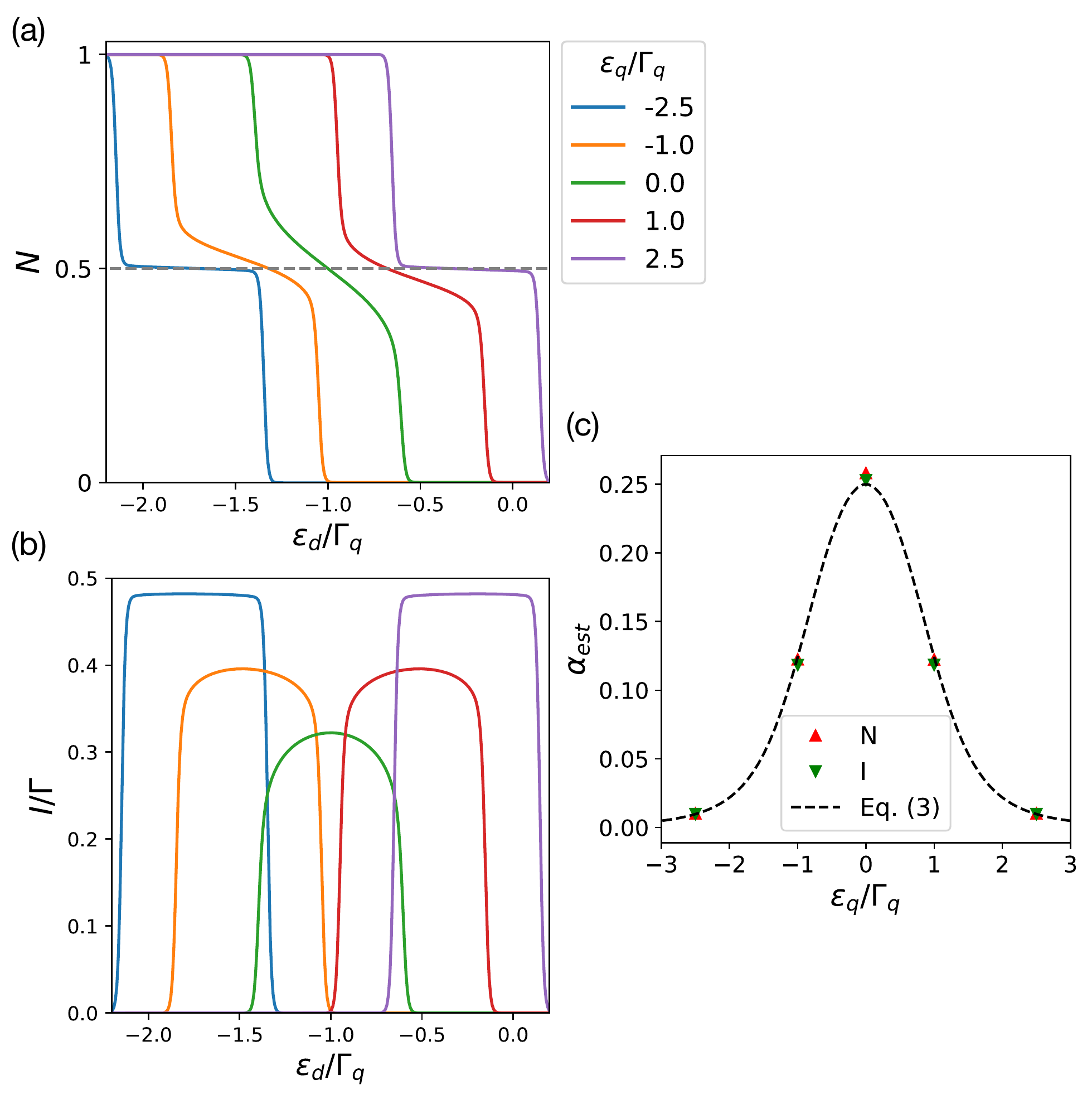}
		\caption{\label{fig:occ_cur_eq}  Additional effects of tuning $\alpha$  by the detector dot level, $\epsilon_q$, on the system dot occupation (a) and steady state current (b). 
			Here $\Gamma_L=\Gamma_R$ and the remaining parameters are (in units of $\Gamma_q$) $\lambda=2,\, T=0.01, \, V=0.8$. Since the Hartree shifts depend on $\epsilon_q$, they are included here.  For $\epsilon_q \ne 0$, the relation $A^+(E)= A^-(-E)$ is not obeyed and this, in the presence of a finite $V$, effectively renormalizes $\Gamma_{L,R}$, leading to $N\ne1/2$ for $\epsilon_d$ in the middle of the bias window and an asymmetric peak for the steady state current. (c) The estimators for $\alpha$ from the occupation using Eq.~\eqref{eq:alpha_suscep_spinless} (red triangles), and from the current using Eq.~\eqref{eq:alpha_cur_spinless} (green inverted triangles) continue to be valid as they agree well the analytical formula using Eq.~\eqref{eq:phase_shift} (black dashed lines).  } 
	\end{figure}
	
	We have so far only considered the case of tuning $\alpha$ by changing $\lambda$ while keeping the detector dot level at $\epsilon_q=0$. In an experiment, however, it is more feasible to tune $\alpha$ by changing $\epsilon_q$. In this section we explore the additional effects of tuning $\alpha$ by $\epsilon_q$.
	
	A trivial effect that occurs when $\epsilon_q$ varies is that the Hartree shift of the $d-$level depends on $\epsilon_q$. This results in a horizontal shift of the occupation and steady state currents versus $\epsilon_d$ as shown respectively in Fig.~\ref{fig:occ_cur_eq}(a), (b). 
	
	At $\epsilon_q=0$, the $A^{\text{\tiny +/}\text{\small -}}$ functions obey the relation $A^+(E)=A^-(-E)$~\cite{sankar2024detector}, but for $\epsilon_q\ne0$ this relation is not satisfied. In the analytical formula for $A^{\pm}(E)$, Eq.~\eqref{eq:apm_analytical}, that is valid for low energies, the violation essentially occurs in the form of unequal coefficients $\mathcal{A}^\pm$. For $\epsilon_d$ near the center of the bias window, since $\Gamma_{L}^{\rm{in}}$ and $\Gamma_{R}^{\rm{out}}$ dominate, it follows from Eq.~\eqref{eq:tunin} and Eq.~\eqref{eq:tunout} that $\Gamma_{L,R}$ effectively get renormalized. Eq.~\eqref{eq:nc} then implies that the dot occupation for $\epsilon_d$ exactly in the middle of the bias window, $N_c$, would change with $\epsilon_q$; in particular for the case where the bare couplings $\Gamma_{L}=\Gamma_{R}$, $N_c=1/2$ occurs only for $\epsilon_q=0$, and there is a systematic shift of $N_c$ away from it, as can be seen in Fig.~\ref{fig:occ_cur_eq}(a). Likewise the steady state current is also affected by the renormalized $\Gamma_{L,R}$ as $\epsilon_q$ changes; in particular for $\Gamma_L=\Gamma_R$, the current peak becomes symmetric only for $\epsilon_q=0$, as shown in Fig.~\ref{fig:occ_cur_eq}(b).
	
	The estimators for the AOC exponent $\alpha$ from the occupation, Eq.~\eqref{eq:alpha_suscep_spinless}, and from the steady state current, Eq.~\eqref{eq:alpha_cur_spinless}, are valid for arbitrary $\Gamma_{L,R}$ (within the rate equations validity regime). Thus these estimators remain valid for any $\epsilon_q$ within the validity regime 
	$T \ll V <\Gamma_q$. The only modification for the two formulas is to replace $\epsilon_d=0$ with $\tilde{\epsilon}_d=0$ (dot level in the middle of the bias window), where $\tilde{\epsilon}_d$ refers to the ($\epsilon_q$ dependent) Hartree shifted $\epsilon_d$. A demonstration of the validity of the estimators for different $\epsilon_q$ is shown in Fig.~\ref{fig:occ_cur_eq}(c).

	\section{Capacitive interaction between the QD and its leads
	}
	\label{se:FES}
	In actual experimental setups, capacitive interaction between the system QD and its leads might not be be weak enough to be neglected. Such an interaction  gives rise to the Fermi edge singularity (FES)~\cite{mahan1967excitons,geim1994fermi,abanin2004tunable,frahm2006fermi,krahenmann2017fermi}, which enhances the tunnelling density of states near the Fermi energy of the lead, modifying the relative tunnel-in and tunnel-out rates across the bias window. We now discuss how the AOC signatures  show up in the presence of FES. (Although AOC and FES boil down to the same underlying phenomenon of an abrupt change of ground state of a many-fermion body system, we refer to these two terms selectively to distinguish their origins.) 
	
	To incorporate FES effects we add to the system's Hamiltonian $H_{\rm{{sys}}}$  in Eq.~\eqref{eq:H_sys} an interaction between the QD and its leads, $H_{\rm{{u}}} =H_{{\rm{{u}}},L}+H_{{\rm{{u}}},R}$ where $H_{{\rm{{u}}},i}= u_ic_{i1}^\dagger c_{i1}d^\dagger d$, see Fig.~\ref{fig:fes_occ_cur}. Let us define $n-$dependent lead Hamiltonians $H_{{\rm{lead}},\,n}^{(i)}=  \sum_{\substack{ k}  }\epsilon_{k} c_{ik}^\dagger c_{ik} +H_{u,\,i} $ $(i=L,R)$. In the resulting model when the QD charge $n$ changes, then $H_{{\rm{lead}},\,n}^{(L)}$, $H_{{\rm{lead}},\,n}^{(R)}$ as well as $H_{{\rm{det}},n}$ get modified, leading to a many-body reorganization  in the system's leads and in the detector.  The tunnel rates of the system QD from/to the lead $i$ then become (see appendix \ref{sec:fes_aoc_tun} for a derivation)
	\begin{align}
		\Gamma_{i}^{{\rm{in}}}(\epsilon_d)&=&\gamma_i^2\int_{-\infty}^{\infty}dE\, B_i^+(\mu_i-\epsilon_d-E)A^+(E),\label{eq:tun_in_fes_det}\\
		\Gamma_{i}^{{\rm{out}}}(\epsilon_d)&=& \gamma_i^2\int_{-\infty}^{\infty}dE\, B_i^-(\mu_i-\epsilon_d-E)A^-(E),\label{eq:tun_out_fes_det}
	\end{align}
	where $B_i^{\pm}(E) = \int dt\,  e^{iEt} \tilde{B}_i^{\pm}(t)$ and 
	\begin{align}
		\label{eq:b_plus}
		\tilde{B}_i^+(t)&={\rm{Tr}}\left[\rho_{\rm{{lead,\,0}}}^{(i)} c_{i1}^\dagger(t) e^{itH_{\rm{{lead,\,0}}}^{(i)}}e^{-itH_{\rm{{lead,\,1}}}^{(i)}} c_{i1}(0)\right],\\
		\label{eq:b_minus}
		\tilde{B}_i^-(t)&={\rm{Tr}}\left[\rho_{\rm{{lead,\,1}}}^{(i)} c_{i1}(0)e^{itH_{\rm{{lead,\,0}}}^{(i)}}e^{-itH_{\rm{{lead,\,1}}}^{(i)}}c_{i1}^\dagger(t)\right].
	\end{align}
	Here $\rho_{\rm{{lead,\,n}}}^{(i)} \propto e^{- H_{{\rm{{lead}}},n}^{(i)}/T} $  denotes the density matrix of the system's leads   for QD charge $n$. The interpretation of the Fourier transform of these functions is  parallel to that of $A^{\pm}(E)$: $A^+(E)$ is the probability  that in the tunneling-in process the detector will absorb energy $E$, and  $B_i^+(E)$ is the probability  to remove  an electron with energy $-E$ from the $i$-th lead relative to its chemical potential; thus in $ \Gamma_{i}^{{\rm{in}}}(\epsilon_d)$ we have an electron jumping from a lead state with energy $E+\epsilon_d$ to the $d$ level with energy $\epsilon_d$, transfering energy $E$ to the detector. Likewise, $A^-(E)$ is the probability  that in the tunneling-out process the detector will emit energy $E$, and $B_i^-(E)$ is the probability to have an empty electron state in the $i$-th lead with energy $-E$ relative to its chemical potential.
	
	From now on we assume identical interaction strengths $u_L=u_R=u$. Dropping the subscript $i$ from the $B^\pm$ correlators, their long time behavior can be calculated by an exact diagrammatic expansion analogous to the one performed in Ref.~ \onlinecite{nozieres1969singularities}, giving
	\begin{equation}
		\label{eq:b_long_time}
		\tilde{B}^{\pm}(t)=\frac{\nu_0}{D'^{-\alpha'}}\left(\frac{\pi T}{\pm i  \sinh(\pi T t) }\right)^{1-\alpha'},
	\end{equation}
	where $D'$ is  a high energy cutoff associated with the leads. The FES exponent $\alpha'$ is determined by $u$ via an associated phase-shift $\delta$ according to~\cite{nozieres1969singularities}
	\begin{equation}
		\alpha'=2\frac{\delta}{\pi}-\left(\frac{\delta}{\pi}\right)^2.
	\end{equation}
	Taking the Fourier transform of $B^\pm(t)$ we obtain for $|E|\ll D$,
	\begin{equation}
		\label{eq:bpm_analytical}
		B^{\pm}(E)=\mathcal{B}{\rm{Re}}\left[\frac{e^{\pm i(\pi/2)(1-\alpha')}\Gamma\left(\frac{iE}{2\pi T}+\frac{1-\alpha'}{2}\right)}{\Gamma\left(1+\frac{iE}{2\pi T}-\frac{1-\alpha'}{2}\right)}\right],
	\end{equation}
	where the prefactor $\mathcal{B}\propto \nu_0 \Gamma(\alpha')  T^{-\alpha'}$ up to a non-universal constant. Note that in the limit of vanishing FES, i.e. $\alpha'\to 0$, it follows from Eq.~\eqref{eq:bpm_analytical} that, $ B^{\pm}(E) \propto \nu_0 f(\mp E)$. Plugging this limiting behavior in Eqs.~\eqref{eq:tun_in_fes_det}, \eqref{eq:tun_out_fes_det} we reproduce the tunnel rates in the absence of FES, given in Eqs.~\eqref{eq:tunin}, \eqref{eq:tunout}.
	
	Before presenting results, let us clarify an approximation  in our treatment. The tunneling process involving lead $i$ invokes the FES in lead $i$ and AOC in the detector and in lead $\bar{i}$ (with $\bar{L}=R$ and $\bar{R}=L$). We neglect the later, whose inclusion boils down to a modified AOC exponent $\alpha$. 
	
	More generally, the AOC exponent $\alpha$ can stem from multiple electrostatically coupled metals or gates near the QD. As long as these various metals do not interact with each other, the power-law form of $A^\pm(E)$ remains intact where the value of the effective exponent $\alpha$ is additive. Indeed, in this case the functions $A^\pm(E)$ become products over these different metals. This factorization follows from the definition Eq.~\ref{eq:a_corr_time}.

	\subsection{Results}
\begin{figure}
		\includegraphics[width=\columnwidth]{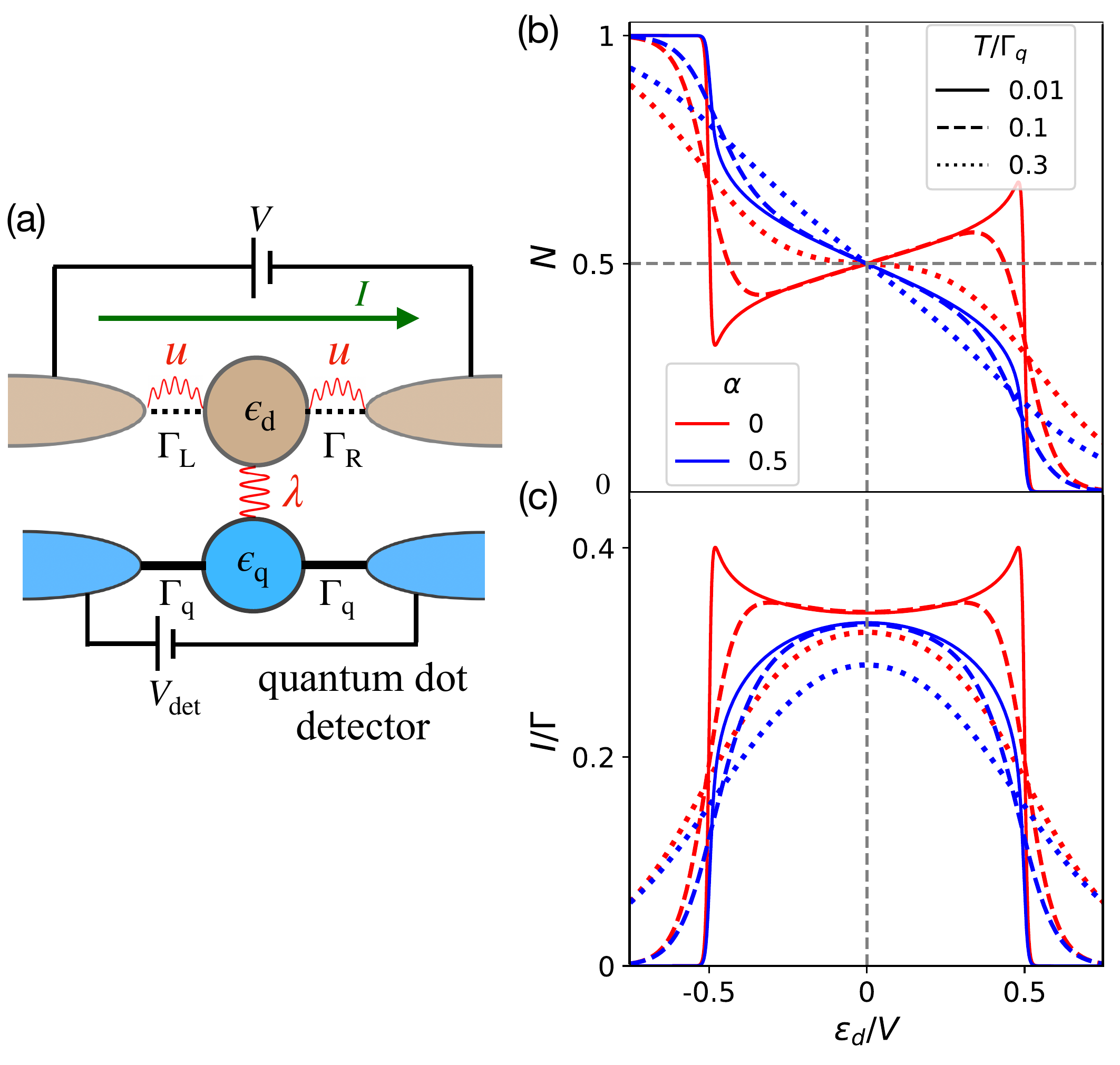}
		\caption{\label{fig:fes_occ_cur} (a) Schematic of the setup where dot-lead interaction, $H_{\rm{u}}$, is included.  The combined effect of FES with leads and detector AOC on (b) the average dot's occupation and (c) the steady state current through the QD.The plots are obtained for the case  $\Gamma_L=\Gamma_R$ and for FES exponent $\alpha'=0.2$ using the respective expressions in Eq.~\eqref{eq:occ_spinless_expr} and Eq.~\eqref{eq:j_spinless_expr} by plugging in the tunnel rates given in Eq.~\eqref{eq:fes_T_tun_in} and Eq.~\eqref{eq:fes_T_tun_out}.} 
	\end{figure}

	Combining the long time behaviors of $\tilde{A}^{\pm}(t)$ and $\tilde{B}^{\pm}(t)$ given respectively by Eqs.~\eqref{eq:a_pm_long_time} and Eq.~\eqref{eq:b_long_time}, we find that the tunnel rates given by Eqs.~\eqref{eq:tun_in_fes_det} ,~\eqref{eq:tun_out_fes_det} are determined  by the difference between the AOC and FES exponents, $\alpha-\alpha'$. In particular, for $|\mu_i-\epsilon_d|\ll D$, we obtain
	\begin{align}
		\Gamma_{i}^{{\rm{in}}}(\epsilon_d)&=\Gamma_i {\rm{Re}}\left[\frac{e^{ i(\pi/2)(1+\alpha-\alpha')}\Gamma\left(\frac{i(\mu_i-\epsilon_d)}{2\pi T}+\frac{1+\alpha-\alpha'}{2}\right)}{\Gamma\left(1+\frac{i(\mu_i-\epsilon_d)}{2\pi T}-\frac{1+\alpha-\alpha'}{2}\right)}\right],\label{eq:fes_T_tun_in}\\
		\Gamma_{i}^{{\rm{out}}}(\epsilon_d)&=\Gamma_i {\rm{Re}}\left[\frac{e^{- i(\pi/2)(1+\alpha-\alpha')}\Gamma\left(\frac{i(\mu_i-\epsilon_d)}{2\pi T}+\frac{1+\alpha-\alpha'}{2}\right)}{\Gamma\left(1+\frac{i(\mu_i-\epsilon_d)}{2\pi T}-\frac{1+\alpha-\alpha'}{2}\right)}\right].\label{eq:fes_T_tun_out}
	\end{align}
	Here we absorbed the non-universal constants into the linewidths $\Gamma_i$. Representative plots of the  QD occupation and steady state current as a function of the QD  energy level in presence of both FES and detector AOC are shown respectivley in Fig.~\ref{fig:fes_occ_cur}(b) and (c).
	
	Let us consider the $T\to 0$ limit. It  follows from Eqs.~\eqref{eq:fes_T_tun_in}, \eqref{eq:fes_T_tun_out} that
	\begin{eqnarray}
		\label{eq:tunin_fes_t0}
		\Gamma_{i}^{{\rm{in}}}(\epsilon_d)&\overset{T\to 0}{=}&\Gamma_i\theta(\mu_i-\epsilon_d) (\mu_i-\epsilon_d)^{\alpha-\alpha'},\\
		\label{eq:tunout_fes_t0}
		\Gamma_{i}^{{\rm{out}}}(\epsilon_d)&\overset{T\to 0}{=}&\Gamma_i\theta(\epsilon_d-\mu_i) (\epsilon_d-\mu_i)^{\alpha-\alpha'}.
	\end{eqnarray}
	Comparing this with Eqs.~\eqref{eq:tunin_t0}, \eqref{eq:tunout_t0} and the subsequent discussion, we reach the conclusion that the relation of AOC exponent $\alpha$ to the charge susceptibility in Eq.~\eqref{eq:alpha_suscep_spinless} and to current derivative in Eq.~\eqref{eq:alpha_cur_spinless} gets simply modified to that of $\alpha-\alpha'$ in presence of FES. Explicitly,
	\begin{align}
		\label{eq:alpha_suscep_fes}
		\alpha-\alpha' &=\frac{-V}{4N_c(1-N_c)}\left(\frac{\partial N}{\partial \epsilon_d}\right)_{\epsilon_d=0},\\
		\label{eq:alpha_cur_fes}
		\alpha-\alpha' &=\frac{\partial \ln[I(\epsilon_d=0)]}{\partial \ln V}.
	\end{align}
	As before, the above relations are valid  in the regime $\{T,V_{\rm{det}}\}\ll V < D$.
	
	\section{AOC signatures in thermal imbalance \label{sec:dif_temp}}
	\begin{figure}
		\includegraphics[width=\columnwidth]{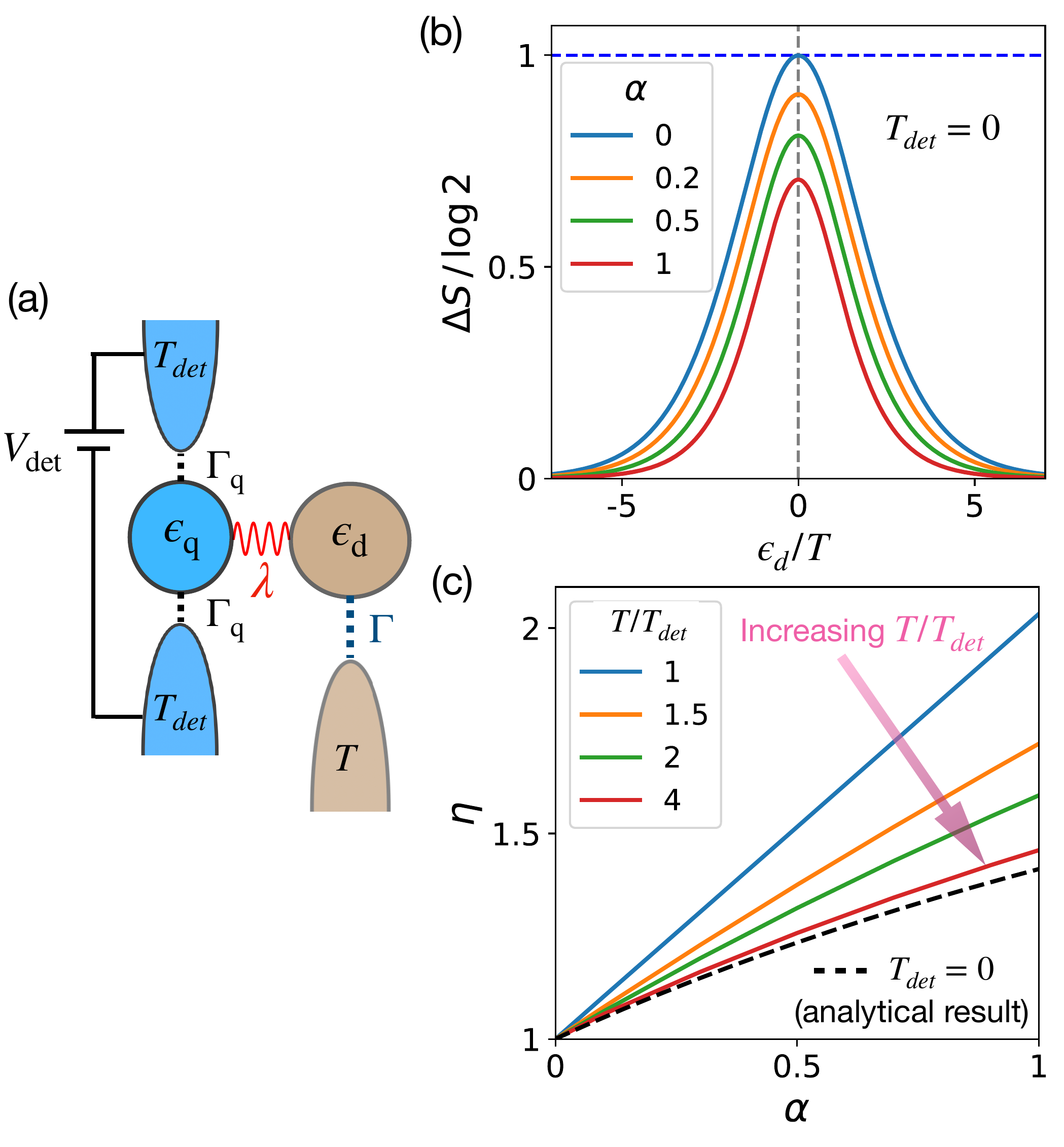}
		\caption{(a) The schematic of the setup with  different temperatures in the detector $(T_{\rm{det}})$ and QD's lead $(T)$. (b) The ``entropy change" of the QD calculated using Eq.~\ref{eq:MR} for  $T_{\rm{det}}=0$. The  entropy peak at $\epsilon_d=0$, $S_{\rm{max}}$, decreases with $\alpha$ reflecting the violation of the thermodynamic Maxwell relation due to thermal imbalance; the ratio $\eta=\log\, 2/S_{\rm{max}}$ provides a direct measure of $\alpha$ and is shown in (c) as dashed lines. (c)  $\eta$ for finite $T_{\rm{det}}$ and $T$ increases monotonically with $\alpha$. Here $T_{\rm{det}}=0.03 D$. $\eta$ converges to the analytic result for $T_{\rm{det}}=0$ as the ratio $T/T_{\rm{det}}$ increases.
		}
		\label{fig:4}
	\end{figure}
	We now consider another non-equilibrium source: the QD leads are at temperature $T$ and charge detector at a different temperature $T_{\rm{det}}$. Here we set $V=V_{\rm{det}}=0$, hence it is sufficient to consider a single QD lead with tunnel coupling $\Gamma = \Gamma_{\rm{L}}+\Gamma_{\rm{R}}$, as depicted in Fig.~\ref{fig:4}(a). As we discuss next, the thermal imbalance affects the QD observables such as $N$ which is no longer given by a Fermi function of $\epsilon_d$. This deviation stems from the AOC exponent $\alpha$ and can be used to extract it.
	
	Let us first consider the case $T_{\rm{det}}=0$ and finite $T$. Plugging  Eq.~\eqref{eq:AT0} in the expression for the tunnel rates in Eq.~\eqref{eq:tunin} and Eq.~\eqref{eq:tunout}, we have
	\begin{equation}
		\Gamma_{\rm{in/out}}=-\Gamma\, \mathcal{A}\, T^\alpha\, {\rm{Li}}_\alpha (-e^{\mp x}),
	\end{equation}
	where $x=\epsilon_d/T$ and $\rm{Li}_\alpha$ is the polylogarithm function of order $\alpha$. The average occupation at steady state follows from Eq.~\eqref{eq:occ_spinless_expr},
	\begin{equation}
		\label{eq:n_x_alpha}
		N(x;\alpha)=\frac{1}{1+ {\rm{Li}_\alpha}(-e^x)/{\rm{Li}_\alpha}(-e^{-x})}.
	\end{equation}
	Note that when $\alpha=0$, this expression reduces to the Fermi function, $N=1/(1+e^x)$. For finite $\alpha$, however, the width of the charge curve reduces. 
	
	The thermal imbalance associated with $\alpha$ can be quantified by the degree of violation of a thermodynamic relation. We consider the violation of the  Maxwell relation, $\partial S/\partial \epsilon_d=-\partial N/\partial T$, where $S$ denotes the thermodynamic entropy of the system. In particular, we consider the entropy change by integrating the Maxwell relation,
	\begin{equation}
		\label{eq:MR}
		\Delta S(\epsilon_d) = \int_{\epsilon_d}^{\infty}\frac{\partial N}{\partial T}|_{\epsilon}d\epsilon.
	\end{equation}
	Such charge detection based entropy measurements of QDs have recently attracted experimental ~\cite{hartman2018direct,child2022entropy,adam2024entropy,Kealhofer_2025,sheekey2026remote,kim2026entropyspectroscopytunabletwosite} and theoretical attention~\cite{ben2015detecting,sela2019detecting,han2022fractional,sankar2023measuring,hurvitz2025metallic,sela2026entropy}. 
	Taking the entropy of the empty QD (i.e $\epsilon_d\to \infty$) to be $0$, we define $
	S_{\rm{max}}=\Delta S (\epsilon_d=0)$. For $\alpha=0$ there is no coupling between the system and the detector, so although they are at different temperatures, the system remains at thermal equilibrium. Then the Maxwell relation is obeyed and we have $S_{\rm{max}}=\log 2$, reflecting the degeneracy of the empty and occupied QD states at $\epsilon_d=0$. But for finite $\alpha$ the Maxwell relation is violated as shown in Fig.~\ref{fig:4}(b). To quantify this nonequilibrium deviation due to the thermal imbalance, we define
	\begin{equation}
		\eta(\alpha)=\frac{\log\, 2}{S_{\rm{max}}(\alpha)}.
	\end{equation} For the case $T_{\rm{det}}=0$ we have
	\begin{equation}
		S_{\rm{max}}(\alpha) \overset{T_{\rm{det}}=0}{=}-\int_0^\infty dx\,x \frac{\partial N (x;\alpha)}{\partial x}=\int_0^\infty dx\, N(x;\alpha).
	\end{equation}
	Plugging in Eq.~\eqref{eq:n_x_alpha}, we then have an analytic expression for $\eta(\alpha)$ at $T_{\rm{det}}=0$, plotted as dashed curve in Fig.~\ref{fig:4}(c).
	
	For finite $T_{\rm{det}}$, the thermal imbalance becomes a function of both $T_{\rm{det}}$ and $\alpha$ and so does $\eta(\alpha)$. Fig.~\ref{fig:4}(c) shows $\eta(\alpha)$ for a fixed $T_{\rm{det}}=0.03\,D$ and different $T$. We find that as the ratio $T/T_{\rm{det}}$ increases, the corresponding $\eta(\alpha)$ converges to the result for $T_{\rm{det}}=0$. Thus, similar to the voltage biased case, a strong non-equilibrium condition due to thermal imbalance allows for an estimation of $\alpha$ in the regime $T_{\rm{det}} \ll T < D$. The proposed scheme to measure $\alpha$ applies for finite $V_{\rm{det}}$ as well, provided the dephasing rate~\cite{sankar2024detector} is small, $\Gamma_\varphi(V_{\rm{det}}) \ll T$.

	\section{Summary} 
	\label{se:con}
	In summary, we discussed simple schemes involving easy-to-measure observables in non-equilibrium conditions, such as the average charge or steady state current, to extract the AOC exponent $\alpha$ for a QD capacitively coupled to a detector. We focused on a non-equilibrium situation emerging from a voltage bias applied on the system QD, but also described effects of a thermal imbalance between the QD's lead and detector temperatures. In both cases, we showed that $\alpha$ can be reliably extracted at strong enough non-equilibrium conditions ($V\gg T$ or $T\gg T_{\rm{det}}$). 
	
	Our predictions in the schemes for case (i)  have been recently implemented experimentally~\cite{grant2026aoc}, showing a very good match with the theory presented here. Such an experimental detection of AOC effects opens the door towards the exploration of more exotic effects such as the quantum phase transitions discussed in Ref.~\onlinecite{goldstein2010population,ma2023identifying}.
	
	\acknowledgements We gratefully acknowledge support from the European Research Council (ERC) under the European Union Horizon 2020 research and innovation programme under grant agreement No. 951541. ARO (W911NF-20-1-0013). Y.M. acknowledges support from ISF Grant No. 359/20 and 737/24. J.F. acknowledges support from the Natural Sciences and Engineering Research Council of Canada; the Canadian Institute for Advanced Research; the Max Planck-UBC-UTokyo Centre for Quantum Materials and the Canada First Research Excellence Fund, Quantum Materials and Future Technologies Program.
	
	\begin{appendix}
		\section{\label{sec:fes_aoc_tun}FES + AOC tunnel rates}
		In this appendix we detail the derivation of the Fermi golden rule based  tunnel rates given in Eq.~\eqref{eq:tun_in_fes_det},~\eqref{eq:tun_out_fes_det}. Let us consider the tunnel-in rate from lead $i$ to the system QD. Let $\{|\phi_k^0\rangle,|\psi_p^0\rangle \}$ respectively denote the eigenbasis of $H^{(i)}_{\rm{lead, 0}},\,H_{\rm{{det}},0}$ and $\{|\phi_l^1\rangle,|\psi_q^1\rangle \}$ respectively denote the eigen basis of $H^{(i)}_{\rm{lead, 1}},\,H_{\rm{{det}},1}$. Let us denote the associated many-body eigen-energies by $E_{k}^0, \,E_{p}^0$ and $E_{l}^1,\,E_{q}^1$, respectively. The initial state of the combined system before the tunnel event is (QD at $n=0$),
		\begin{equation}
			|\psi\rangle_{\rm{{initial}}} = \sum_{k,p}P_kP_p|\phi_k^0\rangle \otimes |\psi_p^0\rangle,
		\end{equation}
		where $P_k=\frac{e^{- E_k^0/T}}{\sum_k e^{-E_k^0/T}}$ and $P_p=\frac{e^{- E_p^0/T}}{\sum_p e^{- E_p^0/T}}$ denote the corresponding thermal probabilities. The Fermi golden rule tunnel-in rate can then be expressed as
		\begin{widetext}
			\begin{eqnarray} \Gamma_{i}^{{\rm{in}}}(\epsilon_d)&=&\gamma_i^2\sum_{\substack{k,l\\ p,q}}\left(P_k\langle \phi_k^0|c_{i1}^\dagger|\phi_l^1\rangle\langle \phi_l^1|c_{i1}|\phi_k^0\rangle\right)\left(P_p\langle \psi_p^0|\psi_q^1\rangle\langle \psi_q^1|\psi_p^0\rangle\right)2\pi \delta(E_k^0+E_p^0-E_l^1-E_q^1+\mu_i-\epsilon_d)\\
				&=& \gamma_i^2 \int dt\,e^{it(\mu_i-\epsilon_d)}\sum_{\substack{k,l\\ p,q}}\left(P_k\langle \phi_k^0|e^{itH^{(i)}_{\rm{{lead}},0}}c_{i1}^\dagger e^{-itH^{(i)}_{\rm{{lead}},1}}|\phi_l^1\rangle\langle \phi_l^1|c_{i1}|\phi_k^0\rangle\right) \left(P_p\langle \psi_p^0|e^{itH_{\rm{{det}},0}} e^{-itH_{\rm{{det}},1}}|\psi_q^1\rangle\langle \psi_q^1|\psi_p^0\rangle\right). \nonumber
			\end{eqnarray}
		\end{widetext}
		Using the identities, $\sum_l |\phi_l^1\rangle\langle\phi_l^1|=1$, $\sum_q |\psi_q^1\rangle\langle\psi_q^1|=1$ along with $\sum_k P_k |\phi_k^0\rangle\langle\phi_k^0|=\rho^{(i)}_{\rm{{lead}},0}$, $\sum_p P_p |\psi_p^0\rangle\langle\psi_p^0|=\rho_{\rm{{det}},0}$, we have
		
		\begin{equation}
			\Gamma_{i}^{{\rm{in}}}(\epsilon_d)=\gamma_i^2\int_{-\infty}^{\infty}dt\, e^{it(\mu_i-\epsilon_d)}\tilde{B}_i^+(t)\tilde{A}^+(t),
		\end{equation}
		where,
		\begin{equation}
			\tilde{B}_i^+(t)={\rm{Tr}}\left[\rho_{\rm{{lead}},\,0}^{(i)} c_{i1}^\dagger(t) e^{itH^{(i)}_{\rm{{lead}},\,0}}e^{-itH_{\rm{{lead,\,1}}}^{(i)}} c_{i1}(0)\right],
		\end{equation}
		with $c_{i1}^\dagger(t)=e^{itH^{(i)}_{\rm{{lead}},\,0}}c_{i1}^\dagger e^{-itH^{(i)}_{\rm{{lead}},\,0}}$. Proceeding similarly, we obtain the expression for the tunnel-out rates
		\begin{equation}
			\Gamma_{i}^{{\rm{out}}}(\epsilon_d)=\gamma_i^2\int_{-\infty}^{\infty}dt\, e^{it(\mu_i-\epsilon_d)}\tilde{B}_i^-(t)\tilde{A}^-(t),
		\end{equation}
		where
		\begin{equation}
			\tilde{B}_i^-(t)={\rm{Tr}}\left[\rho_{\rm{{lead}},\,0}^{(i)} c_{i1}(0) e^{itH_{\rm{{lead}},\,0}^{(i)}}e^{-itH_{\rm{{lead,\,1}}}^{(i)}} c_{i1}^\dagger(t)\right].
		\end{equation}
		
		\section{Spinful generalization}
		\label{sec:spin}
		In the main text we considered a spinless QD, which corresponds to applying a large Zeeman field. Here we comment on the modifications in a spinful model, in the absence of a Zeeman field. 
		
		Adding the spin index, the analysis is only slightly modified, by considering the Anderson model
		\begin{equation}
			\label{eq:h_sys_Anderson}
			\begin{split}
				H_{\rm{sys}} &= \sum_{\sigma=\uparrow,\downarrow}\epsilon_d d_\sigma^\dagger d_\sigma + Ud_\uparrow^\dagger d_\uparrow d_\downarrow^\dagger d_\downarrow \\
				&+\sum_{ik\sigma}\epsilon_k c^\dagger_{i k\sigma}c_{ik\sigma} + \sum_{ik\sigma}\left(\gamma_{i} c^\dagger_{ik\sigma}d_\sigma + {\rm{H. c.}}\right),
			\end{split}
		\end{equation}
		where $U$ is the onsite interaction in the QD.

		Consider a voltage biased QD whose number operator is  $N= \sum_{\sigma}  d^\dagger_\sigma d_\sigma$ with eigenvalues $n=0,1,2$.  As a function of increasing $\epsilon_d$, at $T=V=0$ there are two charge steps: one at $\epsilon_d=-U$ from $n=2$ to $n=1$ and one at $\epsilon_d=0$ from $n=1$ to $n=0$. Let us push $U$ to  large values and focus on the  $n=0 \to 1$ charge step as in Ref.~\onlinecite{child2022entropy}. Upon turning on $V$ this single step splits into two steps at 
		$\pm V/2$ as in Fig.~\ref{fig:1}(b). The only difference in the rate equations is that the tunnel-in rates contain an extra factor of 2 due to spin degeneracy as compared to the tunnel-out rates. As a result, Eq.~(\ref{eq:occ_t0}) obtains the simple modification $\Gamma_L^{(in)}\to 2\Gamma_L^{(in)}$. A similar change occurs in Eq.~(\ref{eq:cur_t0}). Thus our results in Sec.~\ref{sec:biased_QD}  apply with minor modifications in the spinful case provided that $V \ll U$. A similar consideration applies to the scheme discussed for the thermal imbalance case in Sec.~\ref{sec:dif_temp}, and the results there apply with minor modifications in the spinfull case provided that $T,T_{\rm{det}} \ll U$.
	\end{appendix}


\end{document}